# IMPLEMENTATION OF GOLDSCHMIDT'S ALGORITHM WITH HARDWARE REDUCTION

T. Dutta Roy
Illinois Institute of Technology, Chicago, USA

*Abstract*—Division algorithms have been developed to reduce latency and to improve the efficiency of the processors. Floating point division is considered as a high latency operation. This papers looks into one such division algorithm, examines the hardware block diagram and suggests an alternative path which may be cost effective.

I. INTRODUCTION

With the development in the computational complexity of the modern computer applications and the industry wide usage of benchmarks, such as SPECmarks [1], has forced the designers of the general purpose microprocessors to pay particular attention to the implementation of the floating point unit. S. Oberman and Flynn [2] report that in most current processors, division is significantly slower than other operations. Thus faster division algorithms are imperative for recent processors. Further, its very important for the hardware to be as simple as possible and should consume less area. Division algorithms are broadly classified into 2 classes: i. Digit Recurrence Methods [3] and ii. Iterative and Quadratically convergent, considering Functional Iteration, Very High Radix, table lookup and variable latency to be one class.

In this paper, we would examine a division algorithm called Goldschmidt's Algorithm, build hardware block diagram for it, and look for a suitable reuse of the hardware without loosing sync with the global clock. Goldschmidt's Division Algorithm was improved in an excellent manner and examined by [4]. It not only visited division but used this algorithm for Square Root and Square Root reciprocal also.

The basic Idea of this algorithm from the lines developed by [4] can be explained as follows.

Assume Numerator (N) and Denominator (D) to satisfy the constraint $1 \leq N$ and $D < 2$ (considering them as normalized significands of floating point numbers). The basic division method is performed as follows Quotient $Q = N/D$ [4]. The Goldschmidt's algorithm points at finding a sequence $K_1, K_2, K_3 \ldots K_i$ such that the product $r_i = D \cdot K_1 \cdot K_2 \cdot K_3 \cdot K_4 \cdot K_5 \ldots K_i$ approaches 1 as i goes to infinity. Thus we have $q_i = N \cdot K_1 \cdot K_2 \cdot K_3 \cdot K_4 \cdot K_5 \ldots K_i \rightarrow Q$.

$K_1$ is obtained from the lookup table which is an optimal reciprocal table with p-bits in and p+2 bits out that uses D as input and obtains an p+2 bit approximation $K_1$ to 1/D. The input bits and their accuracy considerations were met previously in [4] so we would rather go ahead with the basic idea needed to develop a Hardware architecture and look more into it. The Goldschmidt's algorithm involves 2 steps that can be explained from the hardware block diagram as follows.

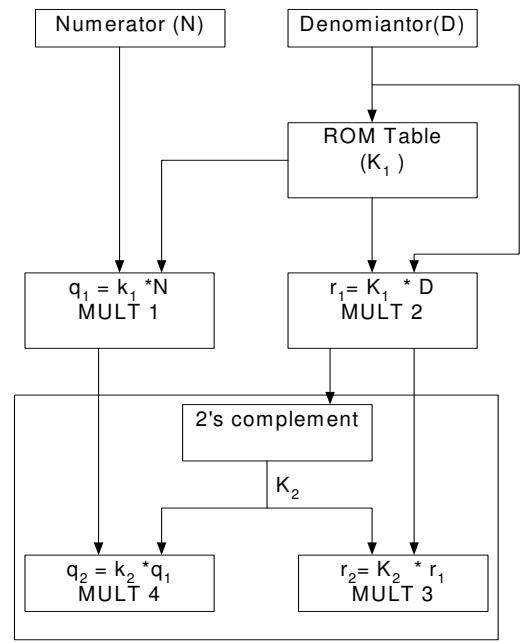

Figure 1. Basic Idea of Goldschmidt's Algorithm

Step 1: The best part of this algorithm is that we have both the numerator and the denominator with us. The denominator is passed through a look-up table in the ROM and the first value of the sequence $K_i$ is obtained. With this value of the sequence we multiply the numerator and the denominator separately and obtain the values $q_1$ and $r_1$. Thus 2 multipliers are used here, MULT 1 and MULT 2 in figure 1.

Step 2: Once we obtain values $q_1$ and $r_1$ the next step involves is the determination of $K_2$, $q_2$ and $r_2$. This can be obtained by taking the 2's complement of $r_1$ to obtain $K_2$. Once $K_2$ is obtained the next step is the determination of $q_2$ and $r_2$, which is just the multiplication of $K_2$ with $q_1$ and $r_1$. This is shown in the figure 1 above.

[4] defines the use of step 2, twice to get the accuracy and does the accuracy calculations for each of the repetitions. The results obtained in the step 2 i.e. $q_2$ and $r_2$ are again passed through the 2's complement block

and multipliers to obtain the values $q_3$ and $r_3$ respectively. This is done once again and $q_4$ gives the result. The figure below shows the remaining part of figure 1.

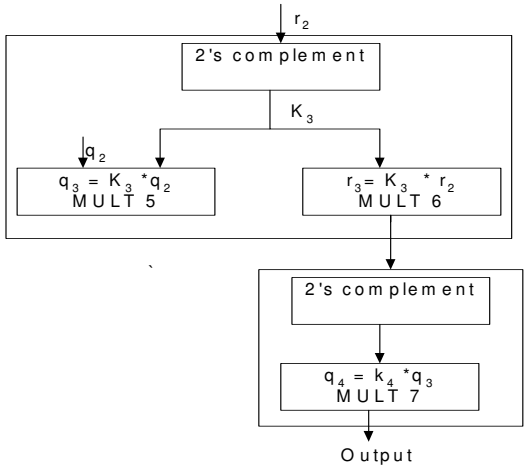

Figure 2: Showing the remaining part of basic Goldschmidt's Algorithm

It implements the above process in pipeline and computes the number of cycles required to obtain the result. The aim of this paper is to reduce the hardware and obtain almost the same accuracy.

## II. CONCEPT

The main idea here is to reuse the multiplier unit again so that the hardware is reduced and then synchronize the logic with the clock such that the total number of cycles required towards the process remains the same. This will make the other variants of the algorithm true and still reduce the area.

Instead of the step 3 in paper [4] which is performing the same multiplication again we decide to add the logic block to the step 2. The output of the step 2 which is $r_2$, is sent to the logic block. The logic block operation is explained as follows.

| LOGIC BLOCK OPERATION | | |
|---|---|---|
| $r_1$ | $r_{2,3...i}$ | O |
| 1 | 0 | $r_1$ |
| 0 | 1 | $r_{2,3...i}$ |
| 1 | 1 | $r_{2,3...i}$ |
| 0 | 0 | 0 |

The logic block output O is $r_1$ when there is no $r_{2,3,..i}$ thus indicating that it's the initial process and then whenever there is $r_{2,3,..i}$ it will give output O as $r_{2,3,..i.}$ Thus giving the idea that we are prioritizing $r_{2,3,..i}$. We can also limit the number of times we need to repeat the multiplication operation and are satisfied with the accuracy by having a unit in built that counts the number of times the value $r_{2,3,..i}$ has been the result of the logic block. This can be predetermined if we are sure of how many bits accuracy we need. Other aspect of consideration is the bit size of the Multiplier X in figure 3 below. This can be fixed with the knowledge of the required output accuracy. However, we realize the fact that multiplication of p x p bit numbers result in (2p-1) bit product. The basic idea remaining making the multiplier n-bit as per accuracy and when the incoming bits is less than n-bits, sensing it and adding leading zeros. This will allow the same multiplier to be used again.

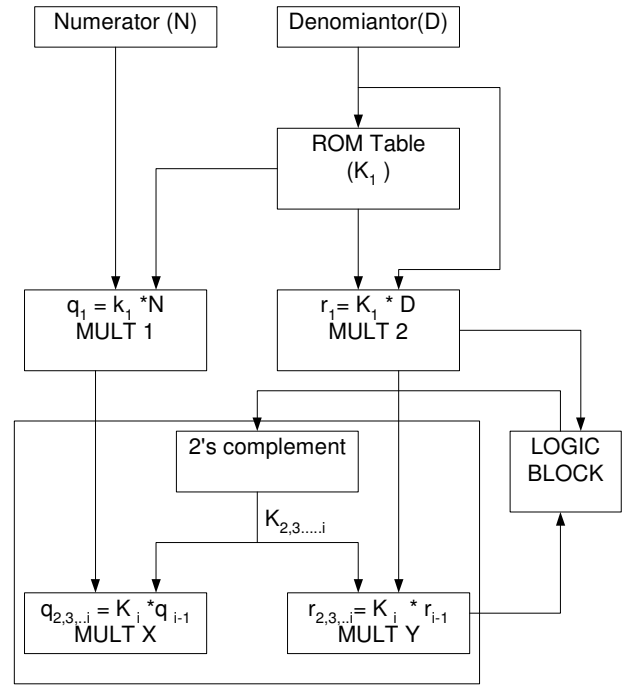

Figure 3. Logic Block Addition

## III. LOGIC BLOCK

The logic block requires $r_1$ to be the input to the 2's complement block for the first time. After that it should discard the value of the $r_1$ until all the values of $r_{2,3,..i}$ are completed.

The idea of creating the logic block is very simple, we have only 2 conditions: 1) When the input is for the first time to the logic block i.e. through $r_1$ then $r_1$ should be the value going into the 2's complement block. 2) Once the value $r_{2,3,..i}$ is ready then it should be value to the 2's complement block. Until all the values of $r_{2,3,..i}$ are passed the output of logic block should remain $r_{2,3,..i.}$ This can be timed using the fact from [4] where we know that a multiplication operation takes 4 cycles to be completed and we need 2 sets of such values, so after 8 cycles the value input should be $r_1$ and not $r_{2,3,..i}$. This can be implemented by using a counter in the logic block which will switch the input from $r_1$ the input first time, to $r_{2,3,..i}$. The input should switch again from $r_{2,3,..i}$ to $r_1$ after the end of predetermined number of cycles as per the accuracy set. Thus we need to implement a counter which will set itself after the first time $r_1$ has passed to $r_{2,3,..i}$ and then again get reset after the predetermined number of cycles are over.

This counter should synchronize with the global clock so that precise operation is done.

IV. COMPARISON WITH THE ORIGINAL METHOD

Using the concept of feed back we can see that the total area consumed can be reduced from the original implementation since the extra multiplier blocks are removed by a feedback unit. However there is a trade off with the speed of operation as pipelining is not done. However, if we assume from [4] that each multiplication operation takes 4 cycles and the used multipliers 1,2, X and Y can be pipelined for the initial value of $r_2$ and $q_2$. The number of cycles taken in both the cases is the same and is 9 cycles with the same factor of accuracy. However, multipliers X and Y can be pipelined amongst themselves. Thus the result of this feedback operation will result into the following

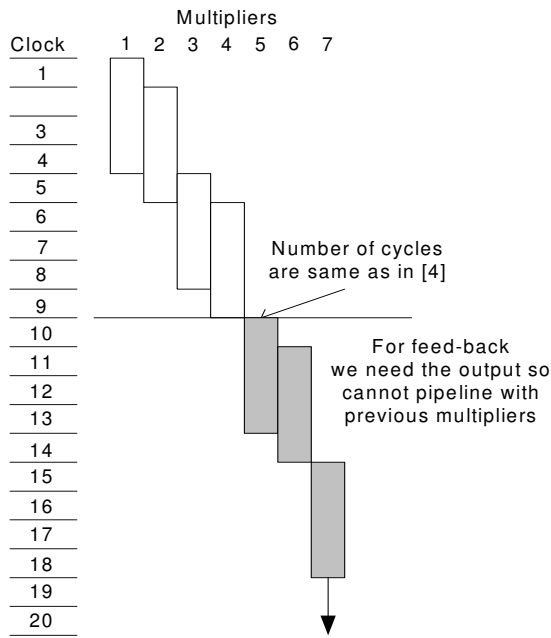

Figure 4: Showing Clock Cycles

The above result is on the basis of the result provided by [4] for its pipelined implementation of Goldschmidt's algorithm. We have developed the concept of feed back to eliminate similar hardware blocks like the 2's complement block and the multiplier and achieved the same accuracy with the trade off of 1 clock cycle for the general case.

A. Variant A[4]

Variant A in [4] remains unaffected as the accuracy result taken from the cycle is used and it perfectly matches the result.

B. Variant B[4]

Here the error term in Variant A is computed and the result is pipelined. However this variation B can be obtained with exactly the same results. Thus making the design compatible with the new idea and area efficient.

V. CONCLUSION

The changes made in the hardware block diagram of the Goldschmidt's algorithm lead to the reduction in the total hardware used even if pipelining was used. Using partial pipelining and feedback, the result with the same accuracy level can be obtained. Further the variants suggested by the paper [4] were not effected at all. The tradeoff between the area and speed was of one clock cycle, where the feedback approach required one clock cycle more, but avoided the use of 3 multipliers and 2 two's complement unit which saves a significant area.